\documentclass{aa}
\usepackage{graphicx}
\usepackage{txfonts}
\usepackage{natbib}
\bibpunct{(}{)}{;}{a}{}{,}
\newcommand\lineone{\ion{Fe}{i}~$\lambda$5250~\AA}

\begin{document}

\headnote{Research Note}
\title{
	Inter-Network magnetic fields 
        observed during\\
	the minimum of the solar cycle
	}	

  \author{
          J. S\'anchez Almeida
	  }

   \offprints{J. S\'anchez Almeida}

   \mail{jos@ll.iac.es}

   \institute{
             Instituto de Astrof\'\i sica de Canarias, 
              E-38205 La Laguna, Tenerife, Spain
	      }

   \date{Received 2 June 2003 / Accepted 29 September 2003}

\abstract{We analyze a time series of high angular resolution magnetograms
of quiet Sun Inter-Network (IN) magnetic fields. These 
magnetograms have a spatial resolution better than 0\farcs5, a noise
of  some 20 G, and they have been obtained at the disk center
during the minimum of the
solar cycle. The IN regions show a typical unsigned flux density of the
order of 15 G. Signals occur, preferentially, in the intergranular
lanes, and the strongest signals trace a network with a scale similar
to the mesogranulation. All these features are consistent with the
IN magnetograms by \citet{dom03a,dom03b}, obtained during the maximum
of the solar cycle. Consequently, the unsigned magnetic flux of the 
structures that give rise to
the IN polarization signals does not seem to
undergo large variations during the solar cycle. 
	}

	\authorrunning{S\'anchez Almeida}
	\titlerunning{Internetwork magnetic fields during solar minimum}

\maketitle	

\section{Rationale}

This note follows up the works
by \citet{dom03a,dom03b}, where we analyzed high spatial
resolution magnetograms of a quiet Sun Inter-Network (IN) region\footnote{The 
IN regions are those photospheric regions appearing as non-magnetic
in routine synoptic magnetograms.
   Magnetic signals in such regions were first reported by
   \citet{liv75} and \citet{smi75}, and they appear
   in the literature with different names, e.g.,
   inner network fields,
   intranetwork fields,
   granular fields,
   turbulent fields, etc.
}.
The IN magnetograms turned out to show much more unsigned 
magnetic flux than the values reported in the 
literature so far. In particular, the flux is larger than the
unsigned flux in the form of active regions, which suggests
the importance of the IN as far as the global magnetic properties
of the Sun is concerned. (See the papers cited above and the 
references therein.)

	Here I analyze a different series of quiet Sun magnetograms
obtained by G. Scharmer\footnote{Quoted to be the series with higher
angular resolution and polarimetric sensitivity obtained 
with the Swedish Vacuum Solar Telescope  and this instrumentation;
Scharmer 2003, private communication.}
during the developmental phases of a magnetograph for the
Swedish Vacuum Solar Telescope \citep[SVST,][]{sch85,sch89}.
They are similar in angular resolution (0\farcs5) and
magnetic sensitivity (20\,G) to those in \citet{dom03a,dom03b} and,
therefore, they offer an independent test for their results.
Moreover, the SVST observations were obtained during the minimum
of the solar cycle (February 1996), as opposed to those of
\citet{dom03a,dom03b}, gathered 6 years later (April 2002). 
This fortunate circumstance allows  us to set constraints on the 
variation of the IN magnetic flux during the solar cycle.

	The dataset and its calibration are discussed in 
Sect.~\ref{observations}. The results are put forward
in Sect.~\ref{results}. The implications of these results
in the context of the previous IN measurements 
are discussed in Sect.~\ref{conclusions}.


\section{Observations and calibrations\label{observations}}

\begin{figure*}
\resizebox{\hsize}{!}{\includegraphics{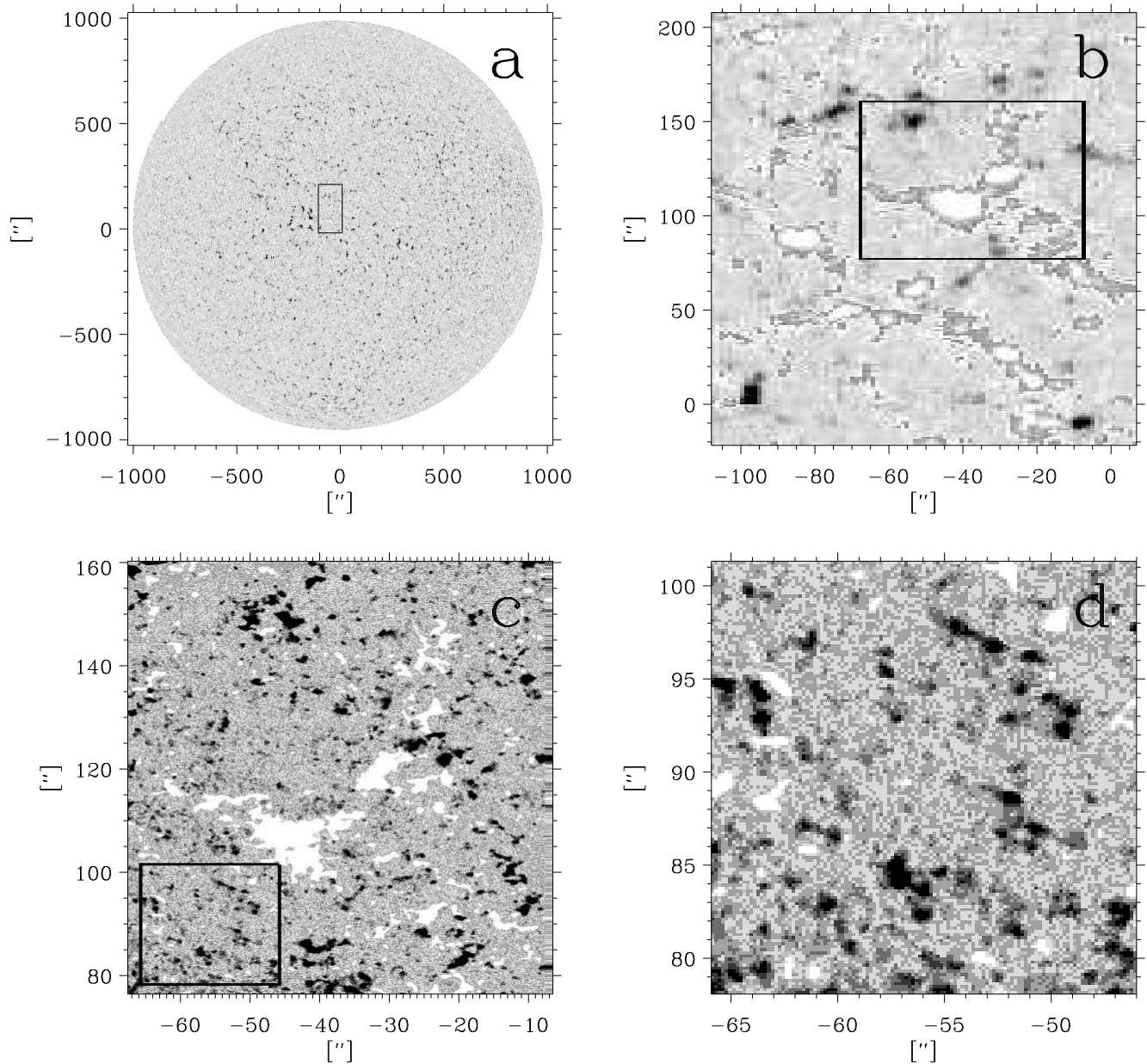}}
\caption{Location of the SVST magnetograms on the solar surface.
{\bf a)} Kitt Peak full disk magnetogram of the same date obtained
some 1/2 hours later than the SVST magnetograms. 
{\bf b)} Blowup of the rectangle in {\bf a}. The new box indicates
the position of the SVST magnetograms. {\bf c)} SVST magnetogram.
{\bf d)} Blowup of  the box in {\bf c}. It corresponds to 
an IN region, as can be deduced by comparison with
the magnetogram in {\bf b}. All spatial scales are in arcsec refereed
to the solar disk center. The scale of grays saturates at $\pm$80\,G.
Note that the aspect ratio 
of the different images is different.
}
\label{whole} 
\end{figure*} 

The quiet Sun region was observed on February 9, 1996, during the
minimum of the solar cycle. A Field-Of-View (FOV) of 
59\arcsec\,$\times$\,81\arcsec\ includes 
several IN regions located very close to the solar disk center (see 
Fig. \ref{whole}; the identification of the FOV within the Kitt Peak
magnetogram was made by \citealp{ste03}). The 0\farcs16 pixel size
slighly undersamples the Airy disk of an ideal telescope with the diameter
of the SVST ($\simeq 50$ cm) at the working wavelength 
($\simeq 5250$ \AA ). The time series
spans some 40 min, with a 1 min cadence. Every snapshot of the
series is made out of 44  images, each one with 210 ms exposure time.  These images were 
selected as those of highest contrast during the 1 min interval of the
cadence. The use of such {\em frame selection} technique 
contributes to the 
good spatial resolution of the resulting magnetograms. In addition,
a tip-tilt mirror corrected for image motion during the individual
exposures. The magnetograph consists of a nematic liquid crystal
variable retarder, followed by a polarizing beam splitter and
a narrow band Lyot filter \citep[see also][]{zha98}. The use of 
the beam splitter ameliorates the  seeing
induced instrumental polarization \citep[see][]{lit87}.
The filter was tuned to select a wing of the magnetically sensitive
line of \ion{Fe}{i} at 5250.2 \AA . It provides only a moderate
spectral resolution (150 m\AA\ at Full Width Half Maximum, FWHM). 
The central wavelength of the observation is unknown, since it
was set by trial and error maximizing the polarization signals
in some test magnetograms.  

\begin{figure}
\resizebox{\hsize}{!}{\includegraphics{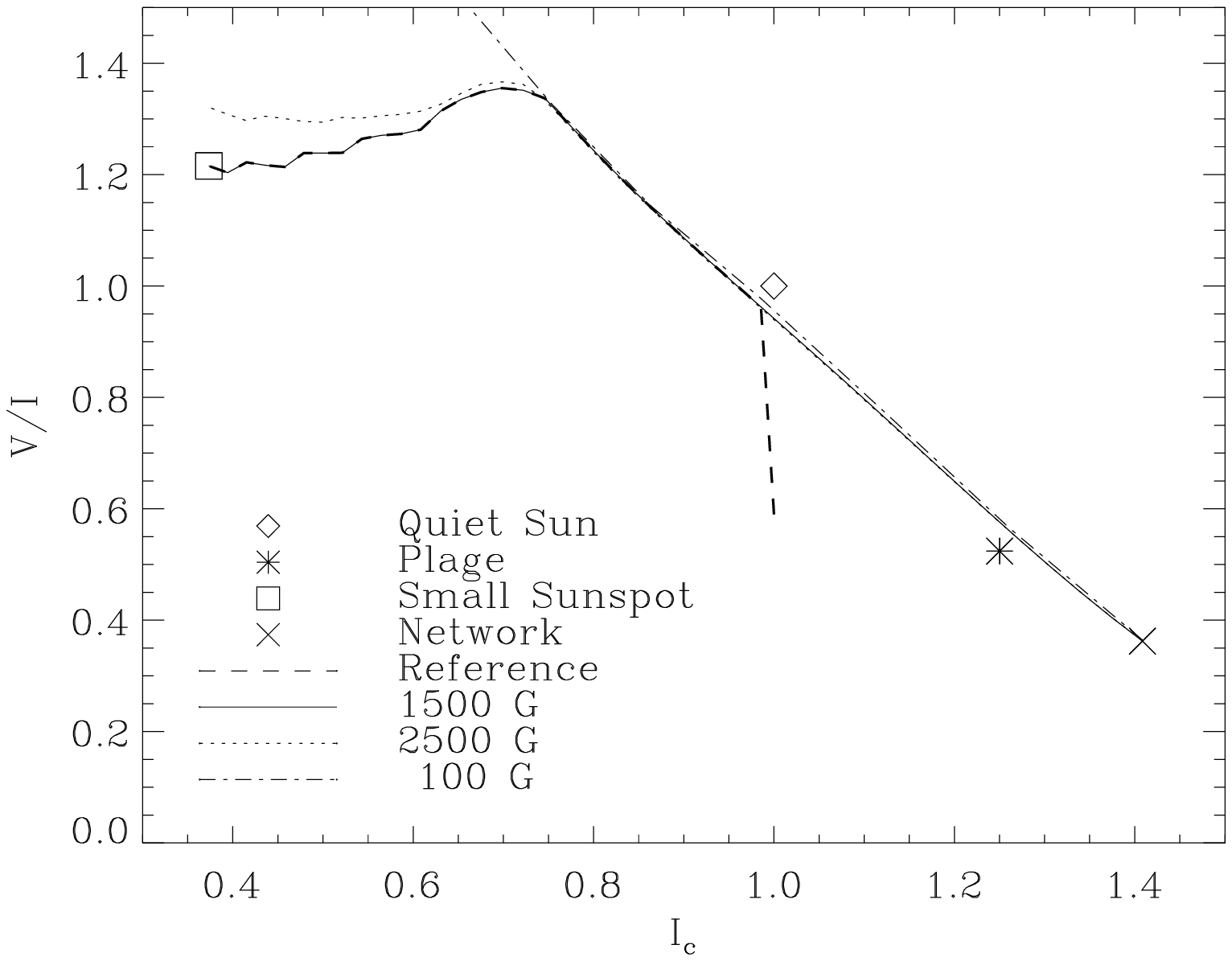}}
\caption{
Synthetic magnetograph signals 
$V/I$ for various model atmospheres
with different thermodynamic
properties, thus leading to different continuum intensity $I_c$. 
The polarization and continuum intensities of \lineone\ are
refereed to the signals in the quiet Sun model atmosphere
by \citet[ the diamond symbol]{mal86}. The figure also includes 
synthetic signals for a model small sunspot \citep[][ the box symbol]{col94},
a  model network \citep[ the times symbol]{sol86},
and a model plage \citep[ the asterisk]{sol86}.
The magnetic field strength is constant with height and identical
in all these four cases (1500~G). 
The microturbulence has been set to 1 km\,s$^{-1}$
whereas the color filter has been assumed to be Gaussian 
with a FWHM of 150~m\AA .
Following the observational procedure, the wavelength of observation
has been chosen as that providing the largest $V$ signal.
The solid line represents a set of model atmospheres whose
thermodynamic conditions are linear combinations
of the two most extreme model atmospheres (network and sunspot).
The dotted line and the dotted-dashed line represent 
the same set of interpolated models except that 
the field strengths have been chosen to be 2500 G and 100 G, 
respectively
(see the inset). 
Finally, the dashed line shows the curve used for calibration. 
It is identical
to the solid line, except that  when  the
continuum intensity was larger than one, then  $I$ from the
quiet Sun synthesis was used to compute $V/I$. This artifice 
tries
to account for the fact that the quiet Sun magnetic structures
producing the observed signals are spatially unresolved
\citep[see, e.g.,][]{dom03a,dom03b}. Then
the observed Stokes $I$ comes from a
mean photosphere,
rather than from the magnetic atmosphere.
}
\label{cal4}
\end{figure}

	The magnetograms were provided uncalibrated, as a degree
of circular polarization $V/I$
at the wavelength of observation. 
They have to be calibrated in
units of flux densities 
(Mx~cm$^{-2}$)
to allow comparison with other observations
	(Since 
	1~Mx~cm$^{-2}=1~$G, it is usual
	to employ G as the flux density unit, which
	is the convention adopted in this note.)
Following the standard procedure, the calibration is based
on the  magnetograph equation which
yields a linear relationship bewteen the  longitudinal
magnetic flux density $B_\mathrm{eff}$ and the 
observed circular polarization 
\citep[e.g.,][]{unn56,lan92},
\begin{equation} 
B_\mathrm{eff}=K~V/I. 
\label{eeqmag} 
\end{equation} 
Calibrating the magnetograms is therefore equivalent to
determining the  constant $K$, which is a non-trivial
step since it  depends on details
of the observational setup (wavelength of observation,
bandpass of the color filter, spectral line, 
thermodynamics of the atmosphere producing the 
polarization,  etc).  
We estimate the calibration constant in the quiet Sun,
$K(\mathrm{quiet})$, from the signal $(V/I)_0$ observed
in a pore existing in the FOV,
\begin{equation}
K(\mathrm{quiet})=\Big[B_0\frac{K(\mathrm{quiet})}{K(\mathrm{pore})}\Big]
	\frac{1}{(V/I)_0}.
	\label{eeqmag3}
\end{equation}
The flux density of the pore $B_0$ is assumed to be known,
\begin{equation}
B_0\simeq 1650\,{\rm G}\pm 250\,{\rm G},
	\label{strength}
\end{equation}
with the error bar accounting for the range of values
found in the literature \citep[ and references therein]{sut98}.
The weakening factor in Eq. (\ref{eeqmag3}), $K(\mathrm{quiet})/K(\mathrm{pore})$,
differs from one, and it has to be estimated. The line 
\lineone\  weakens with increasing temperature, 
and one expects a variation of the calibration
constant between the atmospheres of a cold  pore and the  
magnetic quiet Sun \citep[e.g.,][]{cha68,har69}.
We estimate the weakening factor synthesizing 
Stokes $V$ profiles in various model atmospheres 
with different thermodynamic conditions but
the same flux density. Then the ratio of calibration 
constants between two of such models 
is the inverse of the ratio of magnetograph signals
(see Eq. (\ref{eeqmag})),
\begin{equation}
\frac{K(\mathrm{1})}{K(\mathrm{2})}=\frac{(V/I)_2}{(V/I)_1}.
\end{equation}
The indexes $1$ and $2$ tag the two models.
Using this equation and the dashed curve in Fig. \ref{cal4},
one finds
\begin{equation}
\frac{K(\mathrm{quiet})}{K(\mathrm{pore})}\simeq 1.6\pm 0.4.
	\label{ratio}
\end{equation}
We have assumed the pore 
to be similar to 
a small sunspot, and the magnetic 
quiet Sun thermodynamics to lie in between the
unmagnetized quiet Sun and the network.
This range of possibilities provides
the error bars given in Eq. (\ref{ratio}).
The other curves in Fig. \ref{cal4} show that the
weakening depends very little on the field strength
of the model atmospheres (as far as the 
pore field strength is in the kG regime).
Using Eqs. (\ref{strength}) and  (\ref{ratio}),
\begin{equation}
B_0\frac{K(\mathrm{quiet})}{K(\mathrm{pore})}\simeq 
2600\,{\rm G}\pm 1100\,{\rm G},
	\label{errors}
\end{equation}
which, together with Eqs. (\ref{eeqmag}) and (\ref{eeqmag3}),
render the final calibration,
\begin{equation} 
B_\mathrm{eff}=(2600\,{\rm G} \pm 1100\,{\rm G})~\frac{V/I}{(V/I)_0}.
\label{eeqmag2} 
\end{equation} 
%

The calibration procedure described above has the drawback
of being model dependent. However, it automatically
corrects for several potentially important  systematic effects.
First, the depolarization introduced by the SVST
is believed to be large (e.g., the model Mueller
matrix described by \citealt{san97} shows it to
be as large as 50\%). The use of a ratio of signals,
rather than $V/I$ alone,
cancels it out.\footnote{One 
may also  be concerned by the  contamination
of the Stokes $V$ signals with linear polarization signals.
Although the SVST linear-to-circular polarization crosstalk is 
believed to be large, the solar linear polarization signals
are very small \citep[e.g.,][]{san00}, rendering
a small contamination.}
Second, it accounts for the so-called 
saturation, since the fact that the \lineone\ polarization signals 
are not strictly proportional to the flux density for
kG fields is taken into account by the syntheses.  
Finally, the calibration procedure allows to 
estimate the uncertainties involved in the calibration
procedure, which will be used to 
assess the reliability of the conclusions.

	The noise in the magnetograms was estimated from the difference
between the signals of neighbor pixels, and from signals of the
same pixel obtained in 
consecutive  time steps.
The noise of  neighbor pixels is
independent so, if the true signal in these pixels
is assumed to be the same, then
the difference only bears  noise. Specifically,
the standard deviation
of the difference is $\sqrt{2}$ times the noise in a single pixel. We use the
full series of magnetograms shifted in various
directions to estimate the noise. All of them  
yield similar results; the noise  turns out to be of 
the order of 20\,G, a figure that we adopt  for
the noise level. Note that this value depends on  the calibration
of the magnetograms so it  is 
20$\pm$9\,G, where the error bar just scales the uncertainty
in Eq. (\ref{eeqmag2}).

\begin{figure}
\resizebox{\hsize}{!}{\includegraphics{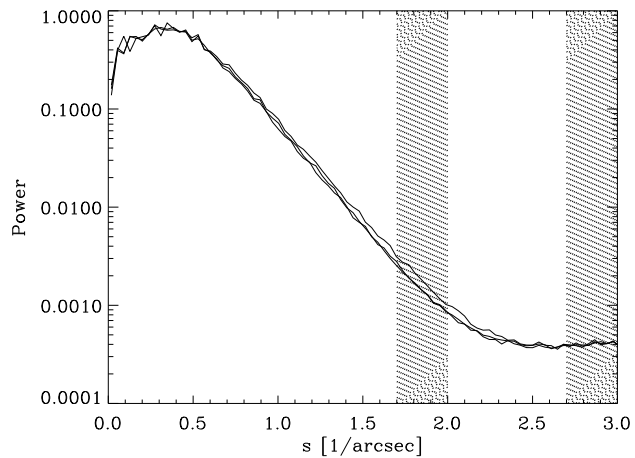}}
\caption{Azimuthally integrated power spectra of the 
three best intensity images in the series. 
The spatial
frequency, $s$, is given in arcsec$^{-1}$. 
The images 
were 
selected as those whose
power in the high-frequency signal bandpass 
(1.7~arcsec$^{-1} < s < 2$~arcsec$^{-1}$) is largest
as refereed to the power in the noise bandpass
(2.7~arcsec$^{-1} < s < 3$~arcsec$^{-1}$). (The
hashed regions of the  plot represent these bandpasses.)
Note that signals exist up to frequencies larger than
2~arcsec$^{-1}$, which corresponds to a period
of 0\farcs5.
The units of the power spectra have been arbitrarily chosen
so that the maximum power is of the order of one. 
}
\label{power}
\end{figure}

In order to characterize the angular resolution of the
observation, we rely on the power spectrum of the 
intensity images of the series.
Figure \ref{power} shows the azimuthally integrated
power spectra of the three best images.
They are the
best images in the sense of having the largest
power in a high frequency bandpass relative to the
power in a  noise bandpass (see the hashed regions 
in Fig. \ref{power}).
It is clear that signals exist up to a frequency
larger than
2 arcsec$^{-1}$,  which corresponds to a period of
0\farcs5.
The contrast of  these best images, i.e., the 
standard deviation over the mean value, is
some 7\%. The best image in the series is chosen in the next
section to quantify the amount of magnetic flux existing
in the IN regions of the magnetograms.
(All good snapshots provide similar fluxes.)

%
\begin{figure}
\resizebox{\hsize}{!}{\includegraphics{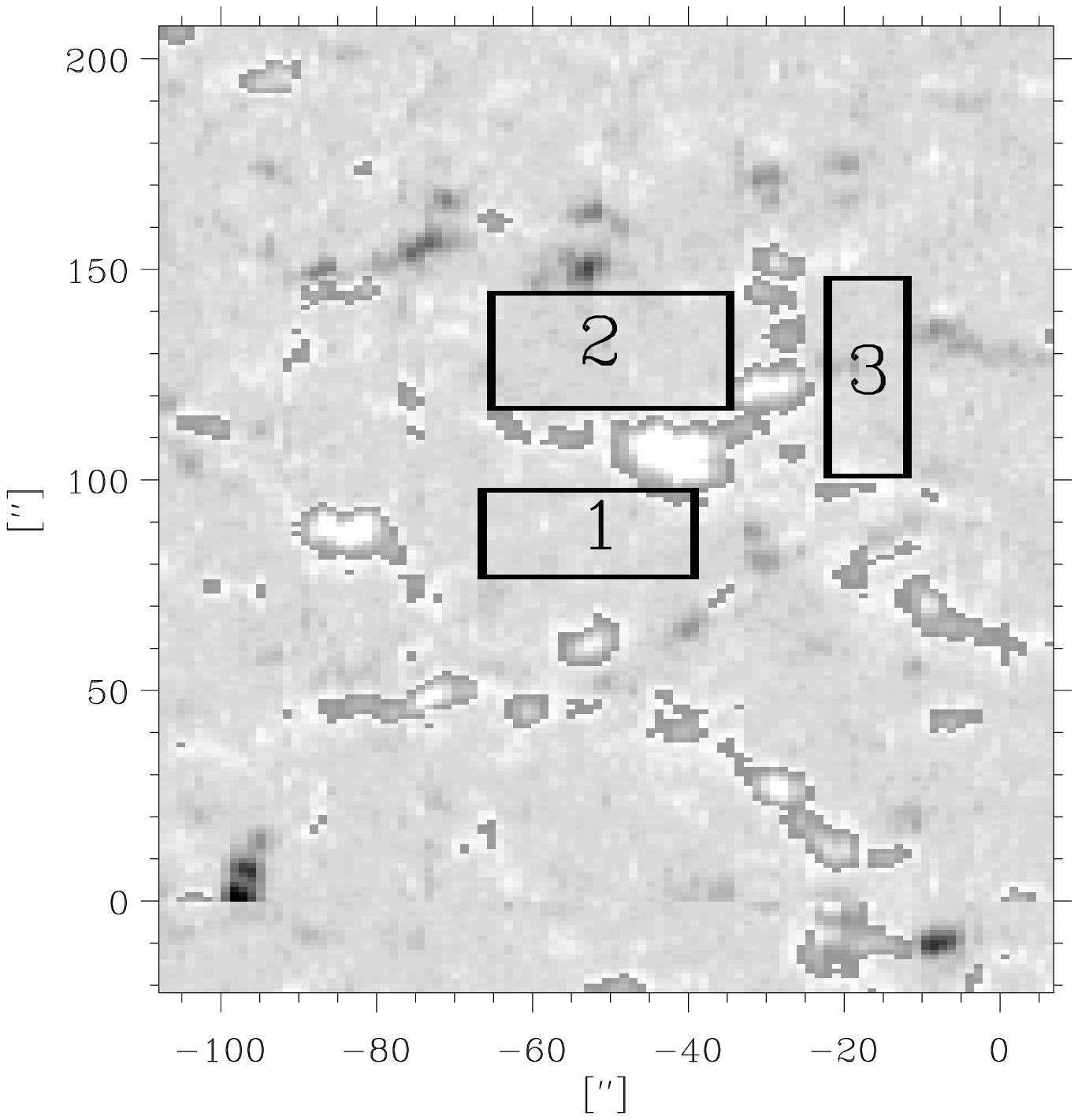}}
\caption{
Kitt peak magnetogram showing the three regions of the SVST
magnetograms used for analysis. Note that they correspond
to IN regions. The FOV is identical to Fig. \ref{whole}b,
but the signals are now scaled between $+$150\,G and $-$150\,G
for clarify.
}
\label{selected} 
\end{figure} 

\section{Results\label{results}}
Following \citet{dom03a,dom03b}, we characterize the 
amount of magnetic signals in the IN regions 
using the mean unsigned flux density, i.e., the
mean value of the calibrated magnetogram
once those signals
below a  threshold have been set to zero.
	(This quantity is defined
	in Eq. (16) of \citealt{dom03b}.) 
As it is worked out in the previous section,
the noise in the magnetograms
is of the order of 20 G. We use a threshold  twice
this value to minimize the contamination
by noise. Three IN regions in the FOV were 
selected for analysis (see Fig. \ref{selected}). The unsigned flux 
densities
of the three regions are given
in Table \ref{table}. All of them show similar
values, which are 
of the order of 15\,G. 
The table also includes the mean signed flux density
(defined as the unsigned flux density but considering the
sign of the signals), and the fraction of FOV covered by magnetic
signals above noise.  The same table contains the unsigned flux density
of the magnetograms analyzed 
in \citet{dom03a,dom03b} once the 40\,G 
threshold used in this paper is taken into account. The
unsigned flux is  some 10\,G and therefore slightly 
smaller than the values that we obtain here.
The same happens with the area covered by the signals, which
is some 17\% instead of the 25\% of the SVST magnetograms.
However, the uncertainty of the calibration 
can easily cope with the difference. Should the calibration constant
be  the lower limit in Eq. (\ref{errors}) (1500 G),
then the mean flux density would drop down to some 8~G.

\begin{figure}
\resizebox{\hsize}{!}{\includegraphics{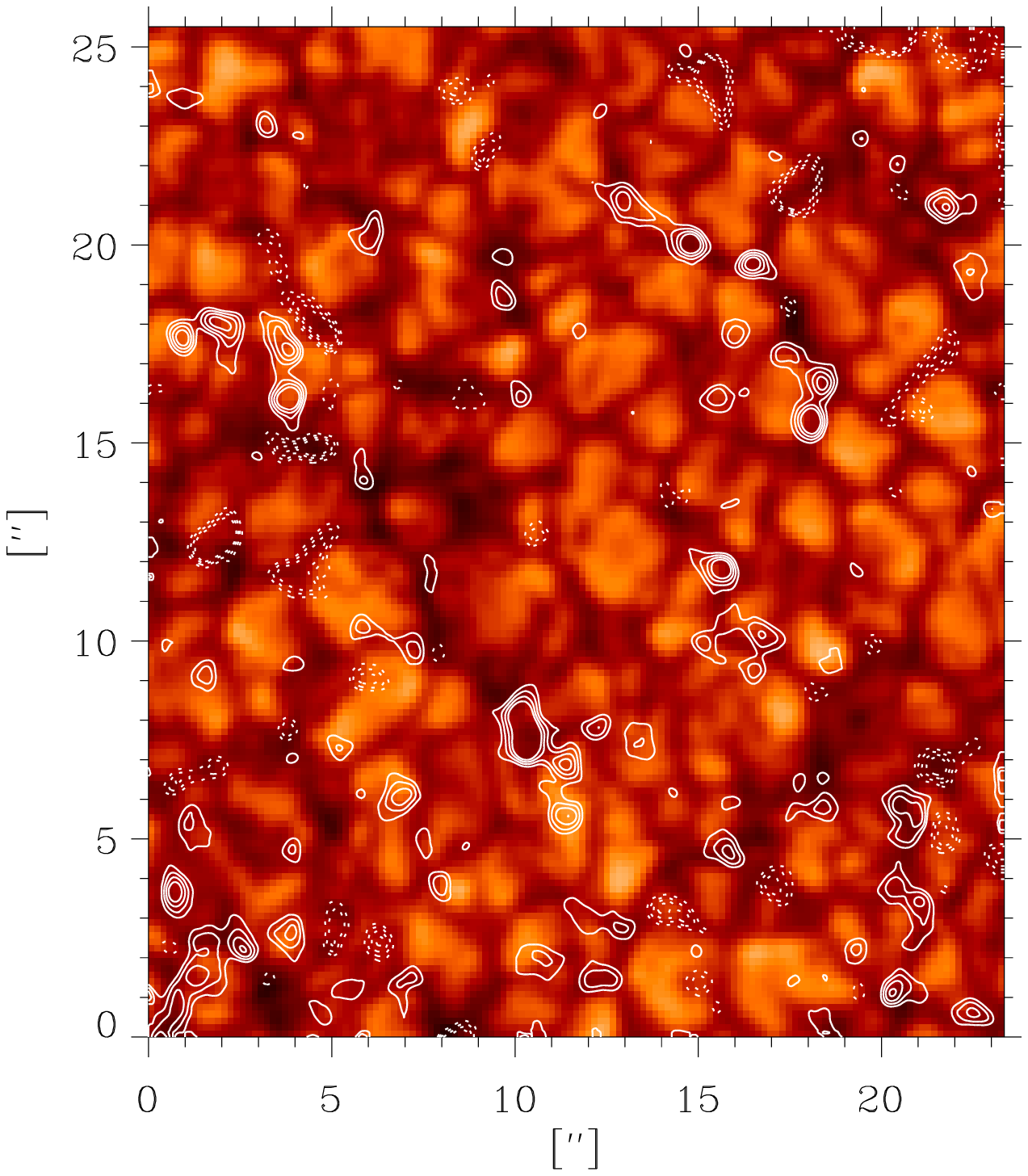}}
\caption{
Intensity image of the IN region \# 1, with the 
contours representing the magnetogram of the same  region.
Note how most of the signals are co-spatial
with intergranular lanes, although this association
is not one-to-one. The contours corresponds to
$\pm$30\,G, $\pm$50\,G, $\pm$70\,G and $\pm$90\,G, with the solid and dotted lines representing
positive and negative polarities, respectively.
}
\label{overlay}
\end{figure}

Another clear result of the  magnetograms of \citeauthor{dom03a}
is the
preference of the signals to appear  on intergranular lanes.
The same tendency is also found  in 
the SVST magnetograms. 
Figure \ref{overlay} shows the intensity image 
of IN region~\#~1 overlaid  with the corresponding
magnetogram represented as contours. 
The magnetic signals show up  on  intergranular lanes,
although not exclusively in there.

\begin{figure*}
\resizebox{\hsize}{!}{\includegraphics{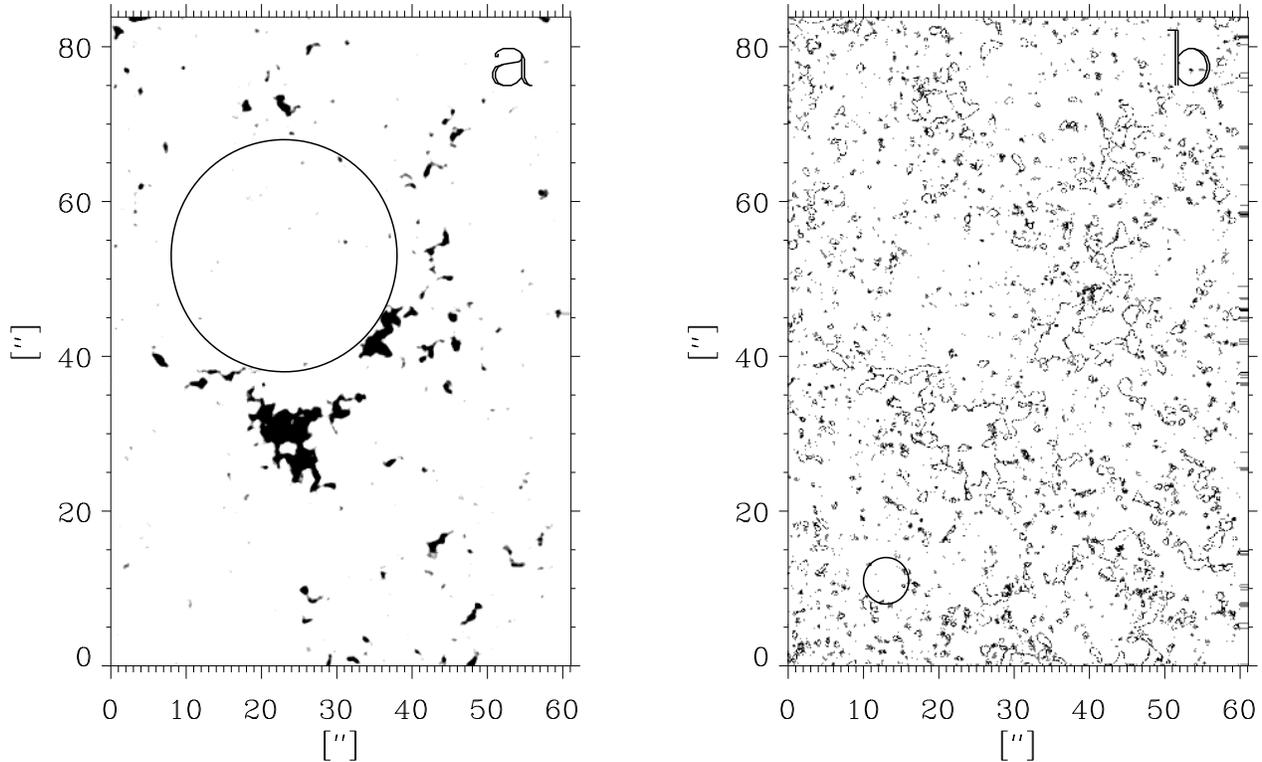}}
\caption{
Absolute value of the signals in the magnetogram for selected
ranges of flux densities. {\bf a)} Strongest
signals in the magnetogram, i.e., those 
whose unsigned flux density is larger than 150\,G. They trace the
network. (The artificial circle, with a diameter of 30\arcsec, 
has the typical size of a network cells.)
{\bf b)} Only strong IN signals are shown (flux densities
between 60\,G and 100\,G). The circle has a diameter of 6\arcsec,
typical of the mesogranular pattern.
}
\label{meso}
\end{figure*}

Yet another result of the works by \citeauthor{dom03a}
is the existence of a web-like pattern 
with the spatial  scale  of the mesogranulation
(say, between 5\arcsec and 10\arcsec ). This
pattern is traced by the largest polarization signals.
Although the pattern is not so clear in the magnetograms studied here,
they contain small regions devoid of strong signals.
Figure \ref{meso} contains two versions of 
the full FOV magnetogram showing signals within
a certain range of flux densities. Figure \ref{meso}a 
represents the strongest signals (larger than 150\,G). It clearly depicts
the supergranulation; the circle in the image has a diameter
of 30\arcsec, characteristic of the network pattern
\citep[e.g.,][]{bec81}.
Figure \ref{meso}b shows the strongest among the IN signals (signals
whose absolute value ranges between 60\,G and 100\,G). Voids with the
mesogranular size are present (the circle on the image,
with a diameter of 6\arcsec, shows a scale typical 
of the mesogranular pattern; see, e.g.,
\citealt{nov81,deu89}). 

%
%
   \begin{table*}
      \caption[]{Mean flux densities in the magnetograms.}
      \label{table}
     $$ 
         \begin{tabular}{ccccc}
            \hline
            \noalign{\smallskip}
	    & Unsigned Flux$^{\mathrm{a}}$ & Signed Flux & 
	    Surface Coverage & Threshold$^{\mathrm{a}}$ \\
	    & [G] & [G] & \%& [G]\\ 
            \noalign{\smallskip}
            \hline
            \noalign{\smallskip}
	    IN Region \# 1$^{\mathrm{b}}$ & 15$\pm$6 & -5 & 28& 40$\pm$17\\
	    IN Region \# 2$^{\mathrm{b}}$ & 13$\pm$6 & -1 & 22& 40$\pm$17\\
	    IN Region \# 3$^{\mathrm{b}}$ & 17$\pm$7 & -3 & 26& 40$\pm$17\\
	    full FOV & 30$\pm$13 & 11 & 30& 40$\pm$17\\
	    reference$^{\mathrm{c}}$ &10& &17& 40\\
	    reference$^{\mathrm{c}}$ &14& &28& 30\\
	    Kitt Peak full FOV$^{\mathrm{d}}$ & 9 & 8 & 8& 40\\
            \noalign{\smallskip}
            \hline
         \end{tabular}
     $$ 
\begin{list}{}{}
\item[$^{\mathrm{a}}$] The error bars account for the uncertainties
in the calibration given in Eq. (\ref{eeqmag2}).
\item[$^{\mathrm{b}}$] Defined in Fig. \ref{selected}.
\item[$^{\mathrm{c}}$] From the \ion{Fe}{i}~6301~\AA\ magnetogram
	in \citet{dom03a,dom03b}.
\item[$^{\mathrm{d}}$] For the box shown in Fig. \ref{whole}b.
\end{list}
   \end{table*}

\section{Conclusions\label{conclusions}}

We estimate that the  unsigned IN flux density at the disk center
is of
the order of 15$\pm$6\,G. Some 25\% of the IN regions is covered by 
signals above 40\,G.
(See Table \ref{table} for a summary of results.) 
These figures for the unsigned flux density and
area coverage  are, within  uncertainties,
compatible with those obtained
by \citet{dom03a,dom03b} from magnetograms with  similar
angular resolution and a slightly better 
polarimetric sensitivity. The overall agreement has two
main implications. First, this independent 
data set confirms the richness of
IN magnetic features found by \citet{dom03a,dom03b}. 
Second,  the 
SVST data used here was taken during the
solar minimum, as opposed to the data by \citeauthor{dom03a} obtained
at maximum.
Since both  datasets are consistent, we conclude that
the IN flux density does not seem to vary along the cycle by
more than $\pm$40\% (i.e., within the error bars
for the calibration of the present magnetograms).
This narrow margin has to be compared with the signals of active
regions. 
Active region flux
at maximum is more than 10 times larger 
than the signals at minimum
\citep[see, e.g.,][ Ch. 12, Fig. 9]{har93}.
The lack of IN flux variation suggests that 
active regions and IN fields have a different origin.
In particular, it discard that the IN flux results from 
the decay of active regions, since it should be modulated
according to the  sunspot cycle.\footnote{
   The same conclusion is reached 
   from the strikingly different decay rates of 
   active regions and IN fields; see \citet{san03b}.
}
These arguments are borrowed from \citet{hag03}, who 
use them to indicate the need of two uncoupled dynamos
to produce active regions and 
network magnetic fields. 

Obviously, the lack of variation
along the cycle
refers only to the kind of measurements
that we analyze, i.e., 0\farcs5 angular resolution disk center
observations of visible  Zeeman
signals above 40\,G. For example, we have no information
on the variation of the signals with latitude. Similarly, our result
neither contradicts nor supports
claims on the variation of the Hanle signals along the
cycle \citep{fau01}. Hanle signals are tracing weak intrinsic
field strengths which are probably not responsible 
for the polarization signals of visible spectral lines
  \citep[see, e.g.,][]{san00,soc03}.

	Let us finish with a speculative detour.
Stars with convective envelopes 
show emission in UV lines, which 
indicates the presence of hot chromospheres.   
Such emission has two components. One component is associated
with the existence of magnetic fields, since  it
is well correlated with the parameters that characterize
the efficiency of a global stellar dynamo.
The second
component, called {\em basal flux}, is always
present independently of the stellar indexes tracing magnetic
activity. According to the current paradigm,
the origin of the basal flux is non-magnetic, being
the residual heating 
due to dissipation of upward propagating waves. 
(For a full account of the current paradigm, see \citealt{sch95};
see also \citealt{wun02}.)
However, a magnetic field component whose properties
remain constant along the cycle could also produce
a residual chromospheric emission
\citep[see, e.g.,][]{jud98}.
The IN magnetic fields
seems to fulfill this requirement and, therefore, 
they may contribute to the basal flux of 
the Sun and other solar-type stars.


\begin{acknowledgements}
The magnetograms were obtained by G.~Scharmer with the
SVST operated by the Royal Swedish Academy of Sciences
in the Spanish Observatorio del Roque de los Muchachos.
%
%
Thanks are 
due to the L.~Rouppe~van~der~Voort  and D.~Kiselman for help with the data handling.
R. Rutten pointed out the relationship between the results in the
note and the {\em basal flux}.
Comments on the manuscript by 
I.~Dom\'\i nguez~Cerde\~na,
	H.~Hagenaar,
F.~Kneer and 
G.~Scharmer 
were very useful.
The ISO/Kitt Peak data used here are produced cooperatively
by NSF/NOAA, NASA/GSFC, and NOAA/SEL.
%
The work was partly supported by the Spanish {\em Ministerio de Ciencia
y Tecnolog\'\i a},
project AYA2001-1649.
\end{acknowledgements}



\begin{thebibliography}{30}
\expandafter\ifx\csname natexlab\endcsname\relax\def\natexlab#1{#1}\fi

\bibitem[{{Beckers}(1981)}]{bec81}
{Beckers}, J.~M. 1981, in The Sun as a Star, ed. S.~{Jordan}, NASA SP-450
  (Washington: NASA), 11

\bibitem[{{Chapman} \& {Sheeley}(1968)}]{cha68}
{Chapman}, G.~A. \& {Sheeley}, N.~R. 1968, \solphys, 5, 442

\bibitem[{{Collados} {et~al.}(1994){Collados}, {Mart\'\i nez Pillet}, {Ruiz
  Cobo}, {del Toro Iniesta}, \& {V\'azquez}}]{col94}
{Collados}, M., {Mart\'\i nez Pillet}, V., {Ruiz Cobo}, B., {del Toro Iniesta},
  J.~C., \& {V\'azquez}, M. 1994, \aap, 291, 622

\bibitem[{{Deubner}(1989)}]{deu89}
{Deubner}, F. 1989, \aap, 216, 259

\bibitem[{{Dom\'\i nguez Cerde\~na} {et~al.}(2003{\natexlab{a}}){Dom\'\i nguez
  Cerde\~na}, {Kneer}, \& {S\'anchez Almeida}}]{dom03a}
{Dom\'\i nguez Cerde\~na}, I., {Kneer}, F., \& {S\'anchez Almeida}, J.
  2003{\natexlab{a}}, \apjl, 582, L55

\bibitem[{{Dom\'\i nguez Cerde\~na} {et~al.}(2003{\natexlab{b}}){Dom\'\i nguez
  Cerde\~na}, {S\'anchez Almeida}, \& {Kneer}}]{dom03b}
{Dom\'\i nguez Cerde\~na}, I., {S\'anchez Almeida}, J., \& {Kneer}, F.
  2003{\natexlab{b}}, \aap, 407, 741

\bibitem[{{Faurobert} {et~al.}(2001){Faurobert}, {Arnaud}, {Vigneau}, \&
  {Frish}}]{fau01}
{Faurobert}, M., {Arnaud}, J., {Vigneau}, J., \& {Frish}, H. 2001, \aap, 378,
  627

\bibitem[{{Hagenaar} {et~al.}(2003){Hagenaar}, {Schrijver}, \& {Title}}]{hag03}
{Hagenaar}, H.~J., {Schrijver}, C.~J., \& {Title}, A.~M. 2003, \apj, 584, 1107

\bibitem[{{Harvey} \& {Livingston}(1969)}]{har69}
{Harvey}, J. \& {Livingston}, W. 1969, \solphys, 10, 283

\bibitem[{{Harvey-Angle}(1993)}]{har93}
{Harvey-Angle}, K.~L. 1993, PhD thesis, Utrecht University, Utrecht

\bibitem[{{Judge} \& {Carpenter}(1998)}]{jud98}
{Judge}, P.~G. \& {Carpenter}, K.~G. 1998, \apj, 494, 828

\bibitem[{{Landi Degl'Innocenti}(1992)}]{lan92}
{Landi Degl'Innocenti}, E. 1992, in Solar Observations: Techniques and
  Interpretation, ed. F.~{S\'anchez}, M.~{Collados}, \& M.~{V\'azquez}
  (Cambridge: Cambridge University Press), 71

\bibitem[{{Lites}(1987)}]{lit87}
{Lites}, B.~W. 1987, \ao, 26, 3838

\bibitem[{{Livingston} \& {Harvey}(1975)}]{liv75}
{Livingston}, W.~C. \& {Harvey}, J.~W. 1975, \baas, 7, 346

\bibitem[{{Maltby} {et~al.}(1986){Maltby}, {Avrett}, {Carlsson},
  {Kjeldseth-Moe}, {Kurucz}, \& {Loeser}}]{mal86}
{Maltby}, P., {Avrett}, E.~H., {Carlsson}, M., {et~al.} 1986, \apj, 306, 284

\bibitem[{{November} {et~al.}(1981){November}, {Toomre}, {Gebbie}, \&
  {Simon}}]{nov81}
{November}, L.~J., {Toomre}, J., {Gebbie}, K.~B., \& {Simon}, G.~W. 1981,
  \apjl, 245, L123

\bibitem[{{S{\" u}tterlin}(1998)}]{sut98}
{S{\" u}tterlin}, P. 1998, \aap, 333, 305

\bibitem[{{S\'anchez Almeida} {et~al.}(1997){S\'anchez Almeida}, {Collados},
  {Mart\'\i nez Pillet}, {Gonz\'alez Escalera}, {Scharmer}, {Shand}, {Moll},
  {Joven}, {Cruz}, {Diaz}, {Rodriguez}, {Fuentes}, {Jochum}, {Paez},
  {Ronquillo}, {Carranza}, \& {Escudero-Sanz}}]{san97}
{S\'anchez Almeida}, J., {Collados}, M., {Mart\'\i nez Pillet}, V., {et~al.}
  1997, in ASP Conf. Ser., Vol. 118, Advances in the Physics of Sunspots, ed.
  B.~{Schmieder}, J.~C. {del Toro Iniesta}, \& M.~{V\'azquez}, 366

\bibitem[{{S\'anchez Almeida} {et~al.}(2003){S\'anchez Almeida}, {Emonet}, \&
  {Cattaneo}}]{san03b}
{S\'anchez Almeida}, J., {Emonet}, T., \& {Cattaneo}, F. 2003, in ASP Conf.
  Ser., Vol. 307, Solar Polarization 3, ed. J.~{Trujillo-Bueno} \&
  J.~{S\'anchez Almeida} (San Francisco: ASP), in press

\bibitem[{{S\'anchez Almeida} \& {Lites}(2000)}]{san00}
{S\'anchez Almeida}, J. \& {Lites}, B.~W. 2000, \apj, 532, 1215

\bibitem[{{Scharmer}(1989)}]{sch89}
{Scharmer}, G.~B. 1989, in Solar and Stellar Granulation, ed. R.~J. {Rutten} \&
  G.~{Severino}, NATO ASI Ser. 263 (Dordrecht: Kluwer), 161

\bibitem[{{Scharmer} {et~al.}(1985){Scharmer}, {Brown}, {Pettersson}, \&
  {Rehn}}]{sch85}
{Scharmer}, G.~B., {Brown}, D.~S., {Pettersson}, L., \& {Rehn}, J. 1985, \ao,
  24, 2558

\bibitem[{{Schrijver}(1995)}]{sch95}
{Schrijver}, C. 1995, \aapr, 6, 181

\bibitem[{{Smithson}(1975)}]{smi75}
{Smithson}, R.~C. 1975, \baas, 7, 346

\bibitem[{{Socas-Navarro} \& {S\'anchez Almeida}(2003)}]{soc03}
{Socas-Navarro}, H. \& {S\'anchez Almeida}, J. 2003, \apj, 593, 581

\bibitem[{{Solanki}(1986)}]{sol86}
{Solanki}, S.~K. 1986, \aap, 168, 311

\bibitem[{{Stenflo} \& {Holzreuter}(2003)}]{ste03}
{Stenflo}, J.~O. \& {Holzreuter}, R. 2003, in ASP Conf. Ser., Vol. 286, Current
  Theoretical Models and Future High Resolution Solar Observations: Preparing
  for ATST, ed. A.~A. {Pevtsov} \& H.~{Uitenbroek} (San Francisco: Astronomical
  Society of the Pacific), 169

\bibitem[{{Unno}(1956)}]{unn56}
{Unno}, W. 1956, \pasj, 8, 108

\bibitem[{{Wunnenberg} {et~al.}(2002){Wunnenberg}, {Kneer}, \&
  {Hirzberger}}]{wun02}
{Wunnenberg}, M., {Kneer}, F., \& {Hirzberger}, J. 2002, \aap, 395, L51

\bibitem[{{Zhang} {et~al.}(1998){Zhang}, {Lin}, {Wang}, {Wang}, \&
  {Zirin}}]{zha98}
{Zhang}, J., {Lin}, G., {Wang}, J., {Wang}, H., \& {Zirin}, H. 1998, \aap, 338,
  322

\end{thebibliography}

\end{document}